# Pulse-Doppler Signal Processing with Quadrature Compressive Sampling

Chao Liu, Feng Xi, Shengyao Chen, Zhong Liu
Department of Electronic Engineering
Nanjing University of Science and Technology
Nanjing, Jiangsu 210094
People's Republic of China
Email: eezliu@mail.njust.edu.cn

**Abstract**: Quadrature compressive sampling (QuadCS) is a newly introduced sub-Nyquist sampling for acquiring *inphase* and *quadrature* (I/Q) components of radio-frequency signals. For applications to pulse-Doppler radars, the QuadCS outputs can be arranged in 2-dimensional data similar to that by Nyquist sampling. This paper develops a *compressive sampling pulse-Doppler* (CoSaPD) processing scheme from the sub-Nyquist samples. The CoSaPD scheme follows Doppler estimation/detection and range estimation and is conducted on the sub-Nyquist samples without recovering the Nyquist samples. The Doppler estimation is realized through spectrum analyzer as in classic processing. The detection is done on the Doppler bin data. The range estimation is performed through sparse recovery algorithms on the detected targets and thus the computational load is reduced. The detection threshold can be set at a low value for improving detection probability and then the introduced false targets are removed in the range estimation stage through inherent detection characteristic in the recovery algorithms. Simulation results confirm our findings. The CoSaPD scheme with the data at one eighth the Nyquist rate and for SNR above -25dB can achieve performance of the classic processing with Nyquist samples.

# I. Introduction

Pulse-Doppler processing has the ability to detect moving targets in strong clutter environments by exploiting the differential Doppler shifts between the real targets and the clutter, and has acquired wide applications in civil and military air surveillance radars [1, 2]. A common processing scheme using quadrature sampling [3, 4] is shown in Fig.1. The radar echoes are sampled to obtain baseband *inphase* and *quadrature* (denoted by *I* and *Q*) components. After processing the baseband signal through a matched filter and discrete Fourier transform (DFT), the detection threshold is applied for constant false alarm rate (CFAR). The detected target plots are given to the data processor for tracking and other functions.

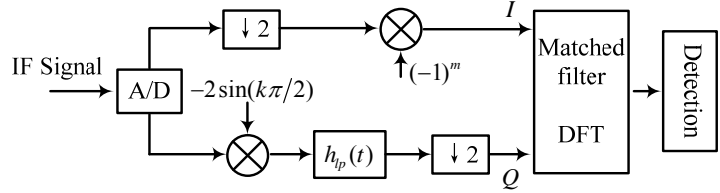

Fig.1: The block schematic of the classic processing.

Assume that the radar echoes are down-converted at intermediate frequency (IF) $f_0$ with bandwidth $B$. Then the Nyquist sampling rate for analog-to-digital conversion (ADC) is given by [5]

$$f_s = \frac{4f_L + 2B}{4l + 1}$$

where $f_L = f_0 - B/2$ and $l$ is a positive integer satisfying $l \leq \lfloor f_L/2B \rfloor$. In the case of wide and ultra-wide band applications, high rate ADC is required and in turn, the intensive processing of high dimensional sequences should be conducted. Currently available ADC technology limits the development of ultra-wideband, high-resolution radar systems.

The newly introduced compressed sensing (CS) [6~8], or compressive sampling, brings us new concepts on the sub-Nyquist data acquisition. The CS theory exploits the sparsity of signals and samples signals closer to their information rate instead of their bandwidth. With high probability, the CS can recover sparse signals from far fewer samples or measurements than the Nyquist samples. The fewer samples lead to a reduced sampling rate and, hence, to a reduced processing load in radar applications. Along with the CS theory, several schemes have been proposed to implement the CS of the analog signals (Analog-to-information conversion, AIC). These include random sampling [9], random demodulation [10], random-modulation pre-integrator [11], segmented compressed sampling [12], Xampling [13], and so on. These schemes have general applicability to signals sparse in time domain, frequency domain or time-frequency domain. Although applicable to the bandpass signals, they do not exploit the characteristics of radar signals and could not directly extract $I$ and $Q$ components from the IF waveform.

Recently, we proposed a quadrature compressive sampling (QuadCS) scheme [14, 15] which merges the CS theory and the digital quadrature sampling. The QuadCS assumes that the echo signals have some sparsity in the waveform-matched dictionary [16] and can directly extract the $I$ and $Q$ components of the bandpass signals. Similar to random demodulation scheme, the QuadCS implements the demodulation through the chipping sequences. However, instead of by the upper frequency of the bandpass signals, the chipping rate is determined by the bandwidth of the bandpass signals.

Therefore, the QuadCS reduces the implementation complexity in comparison with random demodulation scheme.

However, the digital signals obtained through AIC are quite different from that by the Nyquist sampling. The conventional signal processing techniques cannot be directly used for the information extraction. In general, there are two fundamental ways to perform the information processing. One is to do the recovery of the Nyquist sampling signals and process the recovered signals as usual. This kind of processing has the conventional problems of large sampling data and does not utilize the characteristics of the sub-Nyquist data in CS. Another is to process the discrete signals in CS domain. The CS data is small and the direct processing can solve the large data problem in the conventional processing. Signal processing in CS domain, also called as compressive signal processing (CSP), has acquired attention. Some fundamental works, for example, signal detection, parameter estimation, filtering and so on, have been reported [17]. In comparison with conventional techniques, CSP is still in its infancy and much work should be done before practical applications can be developed.

Applications of AIC and CSP to the radar system have been exploited. Some works [18, 19] are towards detection of targets from the sub-Nyquist samples. Others [11, 20~25] are about the extraction of radar target's information (Doppler, range and amplitude) from AIC data. All of these studies demonstrate that the AIC's are effective for radar signal acquisition and the processing load are greatly reduced.

In this paper we discuss the applications of QuadCS to radar and develop a *compressive sampling pulse-Doppler* (CoSaPD) processing scheme. We are mainly concerned with non-fluctuating moving point targets with additive white Gaussian noise (AWGN). The CoSaPD in clutters is also briefly discussed. It is assumed that the radar transmits consecutive pulse trains and target echoes are sampled by the QuadCS. In the range dimension, the QuadCS outputs the discrete data at the sub-Nyquist rate. In the Doppler dimension, the target echoes are sampled at the pulse repetition frequency. Then in a coherent processing interval (CPI), the sampled data can be formulated in a matrix similar to that by the classic sampling (Section IV) [1,2]. Because of the low rate sampling during the intra pulses, the data size is greatly reduced. The targets' information (amplitudes, Doppler frequencies and ranges) are completely contained in the compressive data matrix. Then the target detection and estimation can be performed on the data matrix. The CoSaPD scheme follows the procedures of the Doppler estimation, the target detection and the range estimation. As discussed in Section V, the procedure is irreversible, which is different from the classic processing. Simulations in Section VI show that when the signal-to-noise ratio (SNR) is above -25dB, the CoSaPD scheme at one-eighth the Nyquist rate achieves the performance of classical processing at the Nyquist rate.

The remainder of this paper is organized as follows. In Section II we describe the radar model and the assumptions for our discussion. Section III introduces the fundamentals of the QuadCS system. Section IV describes the proposed processing scheme. The target detection is discussed in Section V, and numerical results are presented in Section VI. We conclude this paper in Section VII.

We denote vectors by boldface lower case letters and matrices by boldface upper case letters. $(\bullet)^{\mathrm{H}}$ denotes the operation conjugate transpose and $(\bullet)^{l}$ denotes the $l$-th column of matrix "$\bullet$". $\mathrm{Re}\{\bullet\}$ and $\mathrm{Im}\{\bullet\}$ represent the real part and the imaginary part of "$\bullet$", respectively. $(\bullet)_{i,j}$ denotes the $i$-th row, $j$-th column element of "$\bullet$".

## II. Radar Model and Problem Statement

In pulse-Doppler signal processing, we usually transmit multiple periodic pulses and perform coherent samples at the range bins to obtain the estimation of target information. We consider the case of $K$ non-fluctuating moving point targets which are sparsely located in the radar's vision and satisfy the stop-and-hop assumption [1]. Assume that the radar transmits a pulse modulated waveform with pulse repetition interval (PRI) $T$ and pulse width $T_b$. Consider the case of $L$ periodic pulses. After downconverting to an intermediate frequency $f_0$, the target echo from the $k$-th target due to the $l$-th transmitting pulse can be described as

$$r_k^l(t) = \rho_k a(t-t_k)\cos[2\pi f_0 t + \phi(t-t_k) + 2\pi f_k^d(l-1)T + \varphi_k], \ t \in [(l-1)T, lT] \quad (1)$$

where $a(t)$ and $\phi(t)$ represent amplitude and phase modulations, respectively; $\rho_k$, $t_k$, $f_k^d$ and $\varphi_k$ are reflecting coefficient, the delay, Doppler frequency and random phase shift of the $k$-th target, respectively. (1) has a bandpass spectrum with center frequency $f_0$ and bandwidth $B$, where $B$ is the bandwidth of transmitting waveform. The received radar echoes due to the $l$-th transmitting pulse is given by

$$r^l(t) = \sum_{k=1}^{K} r_k^l(t) \quad (2)$$

which can be can be expanded as

$$r^l(t) = I^l(t)\cos(2\pi f_0 t) - Q^l(t)\sin(2\pi f_0 t)$$

where $I^l(t)$ and $Q^l(t)$ are called as $I$ and $Q$ components of the signal $r^l(t)$,

$$I^l(t) = \sum_{k=1}^{K} \rho_k a(t-t_k)\cos[\phi(t-t_k) + \varphi'_k]$$
$$Q^l(t) = \sum_{k=1}^{K} \rho_k a(t-t_k)\sin[\phi(t-t_k) + \varphi'_k]$$
(3)

with $\varphi'_k = 2\pi f_k^d (l-1)T + \varphi_k$. Let $\tilde{s}_0(t) = a(t)e^{j\varphi(t)}$ be the complex baseband signal of the radar transmitting signal. Then the complex envelope $\tilde{s}^l(t)$ of $r^l(t)$ is given by

$$\tilde{s}^l(t) = I^l(t) + jQ^l(t) = \sum_{k=1}^{K} \tilde{\rho}_k^l \tilde{s}_0(t-t_k)$$
(4)

where $\tilde{\rho}_k^l = \rho_k \exp\left[j\left(2\pi f_k^d (l-1)T + \varphi_k\right)\right]$.

The target information, $t_k$, $f_k^d$ and $\rho_k$, is completely contained in the complex baseband envelope $\tilde{s}^l(t)$ ($l = 1, 2, \cdots, L$). In the radar signal processing, we usually sample (2) by the quadrature sampling and perform the analysis to obtain the target information, as shown in Fig.1. This paper studies the target estimation through the sub-Nyquist QuadCS data.

To simplify the analysis, we assume that the radar works in unambiguous time-frequency region, *i.e.*, $|f_d| < 1/2T$ and $t_k < T$, and that the target remains in a range bin and keeps constant velocity in a coherent processing interval.

In practical scenarios, the received radar signal inevitably contains noise and clutter in addition to the target echoes[1]. Among various noise sources, thermal noise is nominally dominant. Clutter is often due to echoes from volume or surface scatters [26]. In our study, we assume that the noise is AWGN and the surface clutter is

---
[1] Unintentional electromagnetic interference and intentional jamming are not included in our discussion.

Rayleigh-distributed in amplitude and obeys the two-sided exponential law in Doppler spreading. Then the received radar signal due to the $l$-th transmitting pulse is given by

$$r^l(t) = \sum_{k=1}^{K} r_k^l(t) + n(t) + c(t), \quad t \in [(l-1)T, lT] \tag{5}$$

where $n(t)$ is bandlimited noise with power spectrum density $N_0/2$ and bandwidth $B$, and $c(t)$ is Rayleigh-distributed clutter with average clutter power $\rho_c^2$. We define the received SNR for the $k$-th target as

$$\text{SNR}_k^{IN} = \frac{\frac{1}{T_b}\int_{(l-1)T}^{lT} |r_k^l(t)|^2 dt}{N_0 B} \tag{6}$$

and under the assumption of unit transmitting power, $\text{SNR}_k^{IN} = \frac{|\rho_k|^2}{N_0 B}$. Similarly, the received signal-to-clutter ratio (SCR) can be defined and is given for the $k$-th target by $\text{SCR}_k^{IN} = \frac{|\rho_k|^2}{\rho_c^2}$.

In most part of the paper, we consider the case of target echoes contaminated in the thermal noise $n(t)$. The effects of clutter are analyzed in Section IV and simulated in Section VI.

### III. Fundamentals of Quadrature Compressed Sensing

Now we introduce QuadCS system to perform sub-Nyquist sampling of the received radar signal (5). Different from the system in [14, 15], this work adds the Doppler to the echo model. To simplify the notation, we consider the received signal in a single pulse interval and denote the signal as $r(t)$.

We first consider the case of noise free. CS assumes that the signals $\tilde{s}(t)$ should be sparse in some dictionary. In radar applications, the transmitting waveforms are known in advance. A natural one is the waveform-matched dictionary [16]. For the radar baseband waveform $\tilde{s}_0(t)$ of the bandwidth $B$, the waveform-matched dictionary consists of all the time-delay versions of $\tilde{s}_0(t)$ at integral multiples of $\tau_0 = 1/B$, i.e., $\{\tilde{\psi}_n(t) | \tilde{\psi}_n(t) = \tilde{s}_0(t - n\tau_0), n = 0, 1, \cdots, N-1\}$, where $N = \lceil T/\tau_0 \rceil$ is the size of the dictionary. The dictionary discretizes the observation time $T$ of a pulse with resolution $\tau_0 = 1/B$. This discretization of the time-delay is reasonable, because the time resolution of the bandlimited signal $\tilde{s}_0(t)$ is $1/B$.

Assume that the target delays are at the integral multiples of $\tau_0 = 1/B$, i.e., $t_k \in \{0, \tau_0, \cdots, (N-1)\tau_0\}$. Given the waveform-matched dictionary, the complex envelope $\tilde{s}(t)$ in (4) can be represented as follows:

$$\tilde{s}(t) = \sum_{n=0}^{N-1} \tilde{\rho}_n \tilde{\psi}_n(t) \qquad (7)$$

If there is a target at the delay $t_k$, $\tilde{\rho}_k \neq 0$; otherwise, $\tilde{\rho}_k = 0$. For $K \ll N$, $\tilde{s}(t)$ is said to be $K$-sparse in the waveform-matched dictionary. The sparsity level $K$ exactly equals to the number of the targets.

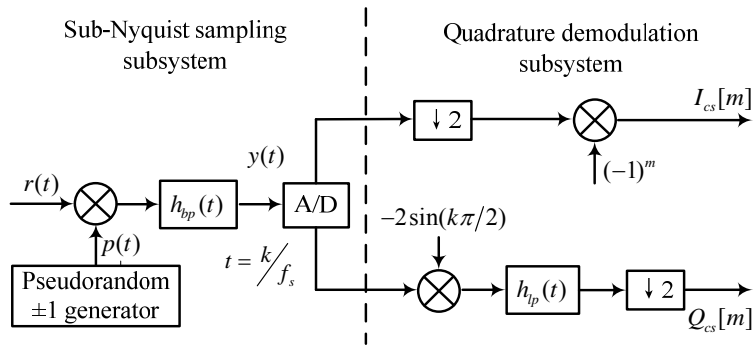

Fig.2: The structure of QuadCS system.

The QuadCS system is shown in Fig.2, which consists of two subsystems: a sub-Nyquist sampling subsystem and a quadrature demodulation subsystem. In the first subsystem, the received radar signal $r(t)$ is modulated by a random chipping sequence $p(t)$ of $\pm 1$'s, which alternates between values at or above the Nyquist rate of the baseband signal. The mixing operation will spread the frequency content of the baseband signal to full spectrum of $p(t)$. Then the mixing output is filtered by a bandpass filter $h_{bp}(t)$ with the center frequency $f_0$ and bandwidth $B_{cs} \ll B$. The filter outputs a *compressive* bandpass signal $y(t)$,

$$y(t) = \int_{-\infty}^{\infty} h_{bp}(\tau) p(t-\tau) r(t-\tau) d\tau = \text{Re}\left\{ \tilde{s}_{cs}(t) e^{j2\pi f_0 t} \right\} \tag{8}$$

where $\tilde{s}_{cs}(t)$ is the *compressive* complex envelope

$$\tilde{s}_{cs}(t) = \int_{-\infty}^{+\infty} h_{bp}(\tau) e^{-j2\pi f_0 \tau} p(t-\tau) \tilde{s}(t-\tau) d\tau \tag{9}$$

with $I_{cs}(t) = \text{Re}\{\tilde{s}_{cs}(t)\}$ and $Q_{cs}(t) = \text{Im}\{\tilde{s}_{cs}(t)\}$ denoting the *compressive I* and *Q* components, respectively. The filter output $y(t)$ is then sampled by a sub-Nyquist ADC to generate a low-rate sequence $y[k]$. The sampling rate is set according to bandpass sampling theorem as $f_{IF}^{cs} = (4f_L + 2B_{cs})/(4l+1)$, where $f_L = f_0 - B_{cs}/2$ and $l$ be a positive integer satisfying $l \leq \lfloor f_L/2B_{cs} \rfloor$.

The second subsystem is to extract digital compressive *I* and *Q* sequences from the sub-Nyquist sampling sequence $y[k]$. Its operation is the same as in classic quadrature sampling [3]. Because of the down-sampling operation, the rate of the digital compressive *I* and *Q* sequences, $I_{cs}[m] = I_{cs}(mT_{cs})$ and $Q_{cs}[m] = Q_{cs}(mT_{cs})$, is half that of $y[k]$, $T_{cs} = 2/f_{IF}^{cs}$. In the observation interval $T$, we obtain $M = \lfloor T/T_{cs} \rfloor$ complex samples $\tilde{s}_{cs}[m] = I_{cs}[m] + jQ_{cs}[m]$ of $\tilde{s}_{cs}(t)$ or $2M$ compressive samples of *I* and *Q* components, which is much less than $2BT$ by the digital quadrature demodulation.

Although the QuadCS system works on analog bandpass signals, its output $s_{cs}[m]$ can be characterized as a linear combination of the sparse coefficient vector $\tilde{\boldsymbol{\rho}} = [\tilde{\rho}_0, \tilde{\rho}_1, \cdots, \tilde{\rho}_{N-1}]^T$. Substituting (7) into (9), we have

$$\tilde{s}_{cs}(t) = \sum_{n=0}^{N-1} \tilde{\rho}_n \int_{-\infty}^{+\infty} h_{bp}(\tau) e^{-j2\pi f_0 \tau} p(t-\tau) \tilde{\psi}_n(t-\tau) d\tau \tag{10}$$

Then

$$\tilde{s}_{cs}[m] = \sum_{n=0}^{N-1} \tilde{\rho}_n \int_{-\infty}^{+\infty} h_{bp}(\tau) e^{-j2\pi f_0 \tau} p(mT_{cs}-\tau) \tilde{\psi}_n(mT_{cs}-\tau) d\tau \tag{11}$$

In the discrete CS framework, we have

$$\tilde{\mathbf{s}}_{cs} = \tilde{\mathbf{M}} \tilde{\boldsymbol{\rho}} \tag{12}$$

where $\tilde{\mathbf{s}}_{cs} = [\tilde{s}_{cs}[0], \cdots, \tilde{s}_{cs}[M-1]]^T$ and $\tilde{\mathbf{M}} = [\tilde{M}_{mn}] \in \mathbb{C}^{M \times N}$ with

$$\tilde{M}_{mn} = \int_{-\infty}^{+\infty} h_{bp}(\tau) e^{-j2\pi f_0 \tau} p(mT_{cs}-\tau) \tilde{\psi}_n(mT_{cs}-\tau) d\tau \tag{13}$$

The recovery of the sparse coefficient vector $\tilde{\boldsymbol{\rho}}$ can be achieved through optimization [27] as

$$\begin{cases} \min \|\tilde{\boldsymbol{\rho}}\|_1 \\ \text{s.t. } \tilde{\mathbf{s}}_{cs} = \tilde{\mathbf{M}} \tilde{\boldsymbol{\rho}} \end{cases} \tag{14}$$

The matrix $\tilde{\mathbf{M}}$ is called the system measurement matrix. For a radar signal having flat spectrum, the matrix $\tilde{\mathbf{M}}$ is approximately column-by-column orthogonal and has nearly same column energy $2T_b B_{cs}^2 / B$ under the assumption of unit transmitting power. The $k$-th target power after the QuadCS system becomes $2|\rho_k|^2 B_{cs} / B$.

When the received signals are contaminated in noise, the QuadCS sampling of

(11) is corrupted by a compressive noise sampling $\tilde{n}_{cs}[m]$, which is obtained by passing the received noise $n(t)$ through the QuadCS system as above. For the additive, white and bandlimited Gaussian noise $n(t)$ with power spectrum density $N_0/2$ and the bandwidth $B$, the compressive noise sampling $\tilde{n}_{cs}[m]$ is an independently, identically distributed (*i.i.d*) complex Gaussian process with zero-mean and variance $2N_0 B_{cs}$. Then output SNR of the QuadCS system from the $k$-th target, the *compressive* $\text{SNR}_k^{CS}$, keeps intact.

In the noisy case, (12) is given as

$$\tilde{\mathbf{s}}_{cs} = \tilde{\mathbf{M}}\tilde{\boldsymbol{\rho}} + \tilde{\mathbf{n}}_{cs} \tag{15}$$

where $\tilde{\mathbf{n}}_{cs} = \left[\tilde{n}_{cs}[0], \cdots, \tilde{n}_{cs}[M-1]\right]^T$. The representation is reasonable because the QuadCS is a linear system. The reconstruction of the sparse coefficient vector $\tilde{\boldsymbol{\rho}}$ in noise case is to solve [28],

$$\min_{\tilde{\boldsymbol{\rho}}} \ \frac{1}{2}\left\|\tilde{\mathbf{s}}_{cs} - \tilde{\mathbf{M}}\tilde{\boldsymbol{\rho}}\right\|_2^2 + \lambda \|\tilde{\boldsymbol{\rho}}\|_1 \tag{16}$$

where $\lambda > 0$ is the regularization parameter which is used to establish the cost of complexity relative to the least-squares error $0.5\left\|\tilde{\mathbf{s}}_{cs} - \tilde{\mathbf{M}}\tilde{\boldsymbol{\rho}}\right\|_2^2$.

There are a wide variety of approaches to solve (14) and (16), including the greedy iteration algorithms [29, 30] and convex optimization algorithms [28, 31] (see [32] for a review). In the simulation study, we use basis pursuit denoising (BPDN) [28] to find the sparse vector $\tilde{\boldsymbol{\rho}}$.

## IV. Pulse-Doppler Processing in QuadCS Domain

This section discusses the extraction of the target information, ranges and Doppler frequencies, with the data given by (15).

Consider a coherent processing interval consisting of $L$ periodic pulses and denote the output of the QuadCS system from the $l$-th echo as

$$\tilde{\mathbf{s}}_{cs}^{l} = \tilde{\mathbf{M}}\tilde{\boldsymbol{\rho}}^{l} + \tilde{\mathbf{n}}_{cs}^{l} \qquad (17)$$

Define $\tilde{\mathbf{S}}_{cs} = [\tilde{\mathbf{s}}_{cs}^{1}, \tilde{\mathbf{s}}_{cs}^{2}, \cdots, \tilde{\mathbf{s}}_{cs}^{L}]$, $\tilde{\boldsymbol{\Theta}} = [\tilde{\boldsymbol{\rho}}^{1}, \tilde{\boldsymbol{\rho}}^{2}, \cdots, \tilde{\boldsymbol{\rho}}^{L}]$ and $\tilde{\mathbf{N}}_{cs} = [\tilde{\mathbf{n}}_{cs}^{1}, \tilde{\mathbf{n}}_{cs}^{2}, \cdots, \tilde{\mathbf{n}}_{cs}^{L}]$. Then the sampling data of the $L$ consecutive echoes can be expressed in a matrix form as

$$\tilde{\mathbf{S}}_{cs} = \tilde{\mathbf{M}}\tilde{\boldsymbol{\Theta}} + \tilde{\mathbf{N}}_{cs} \qquad (18)$$

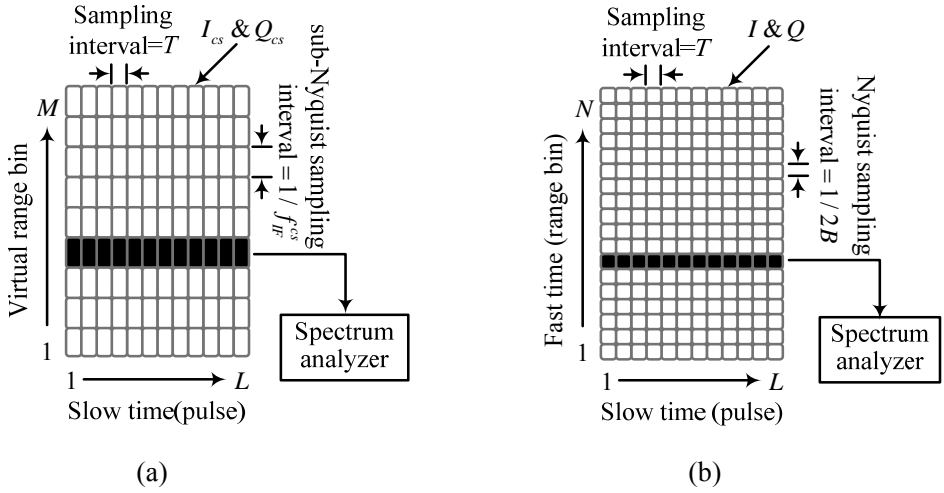

Fig.3: Notional 2-dimensional data matrix generated by the QuadCS system (a) and classic sampling (b).

Fig.3(a) shows 2-dimensional data matrix generated by the QuadCS system. To make a comparison, Fig.3(b) gives the 2-dimentional data matrix by the classic sampling [1,2]. It can be seen that the sub-Nyquist samples by the QuadCS system correspond to the fast time samples (range bins) in the classic sampling. For convenience, we also call the sub-Nyquist samples as the *virtual range bins*. The samples in each column are successive samples of the echoes from a single pulse, *i.e.*, successive virtual range bins. Each element of a column is one complex number, representing the real and imaginary ($I_{cs}$ and $Q_{cs}$) components of one virtual range bin. Consequently, each row represents a series of measurements from the same virtual range bin over successive pulses. Because of the sub-Nyquist rate in the range dimension, the data size in range dimension is greatly smaller than that by the classic fast time sampling.

As seen from (18), the target information is completely included in the $N \times L$ data matrix $\tilde{\mathbf{\Theta}}$. In fact, the data matrix (18) will degenerate to the classic data matrix when $\tilde{\mathbf{M}} = \mathbf{I}_N$. Then if we would obtain $\tilde{\mathbf{\Theta}}$, we could estimate the target information as usual. However, the data which is available is actually an $M \times L$ underdetermined data matrix $\tilde{\mathbf{S}}_{cs}$ because of $M \ll N$. It is impossible to obtain the target information directly. In ideal case, each column of $\tilde{\mathbf{\Theta}}$ is sparse because the number of targets is much less than that of the range bins or the dictionary size. We can firstly obtain the sparse estimates of $\tilde{\mathbf{\Theta}}$ in (18) by solving $l_1$-norm optimization

$$\min_{\tilde{\boldsymbol{\rho}}^l} \frac{1}{2} \left\| \tilde{\mathbf{s}}_{cs}^l - \tilde{\mathbf{M}} \tilde{\boldsymbol{\rho}}^l \right\|_2^2 + \lambda \left\| \tilde{\boldsymbol{\rho}}^l \right\|_1 \quad l = 1, 2, \cdots, L \tag{19}$$

Then the target information can be estimated from row DFT of the estimated $\tilde{\mathbf{\Theta}}$. In practice, due to the influence of noise and clutters, we can hardly obtain the exact information of the targets and may derive false targets. In addition, it takes large computational load by directly solving (19), which is not feasible for real-time processing.

It is seen that each row of the data matrix $\tilde{\mathbf{S}}_{cs}$ represents a series of measurements over successive pulses from the same virtual range bin. Then the target Doppler frequencies can be estimated by the spectral analysis of the slow-time data for each virtual range bin. A simple technique is to conduct the DFT. Denote $\mathcal{F}(\bullet)$ as the DFT of "•" in row vectors. We have

$$\begin{aligned}\mathcal{F}(\tilde{\mathbf{S}}_{cs}) &= \mathcal{F}(\tilde{\mathbf{M}}\tilde{\mathbf{\Theta}}) + \mathcal{F}(\tilde{\mathbf{N}}_{cs}) \\ &= \tilde{\mathbf{M}}\mathcal{F}(\tilde{\mathbf{\Theta}}) + \mathcal{F}(\tilde{\mathbf{N}}_{cs})\end{aligned} \quad (20)$$

Each element of the matrix $\mathcal{F}(\tilde{\mathbf{S}}_{cs})$ is a Doppler spectrum sample, corresponding to the virtual range bin and the frequency bin. Then each Doppler spectrum sample can be detected to determine whether the target is present at the virtual range bin and the Doppler bin.

The DFT plays a role of a matched filter for slow-time samples in the assumed scenarios. After DFT processing, the $k$-th target power becomes $2L^2|\rho_k|^2 B_{cs}/B$ and noise variance becomes $2LN_0 B_{cs}$. Then the SNR after DFT from the $k$-th target, $\text{SNR}_k^{DFT}$, is enhanced to $L$ times the received $\text{SNR}_k^{IN}$. In fact, from the point view of the target detection, we can further improve the detection performance by performing the matched filter for the sub-Nyquist samples, which corresponds to the matching filtering at each Doppler bin. The details are discussed in the next section.

However, the detection process only detects the existence of targets in the specific Doppler bins. We cannot derive the number of the targets and the corresponding ranges. Note that the sparsity of $\mathcal{F}(\tilde{\mathbf{\Theta}})$ can be greatly enhanced even for practically non-sparse $\tilde{\mathbf{\Theta}}$ after the DFT processing. Then for the under-determined data $\tilde{\mathbf{S}}_{cs}$, we may obtain the estimates of the number and ranges of the targets by solving the sparse solution of each column of (20). But the estimation methods will take large computational load because we have to perform the sparse estimates for each column of (20). Since we have detected the targets from the Doppler spectrum samples, we just need to perform the estimation of the target ranges for the detected targets in the specific Doppler bin. In this way, we can greatly reduce the computational load because the estimation of the target ranges only takes place in the corresponding columns.

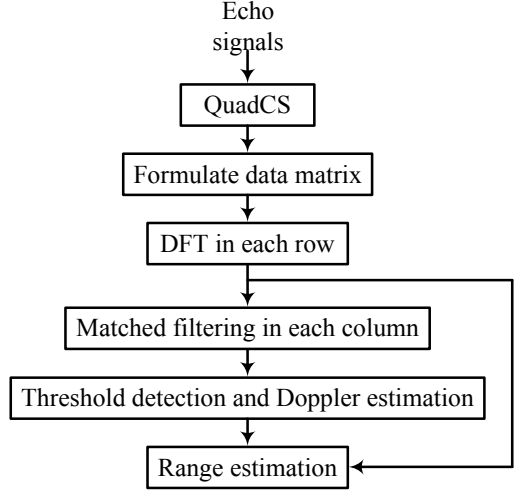

Fig.4: The block schematic of the CoSaPD processing.

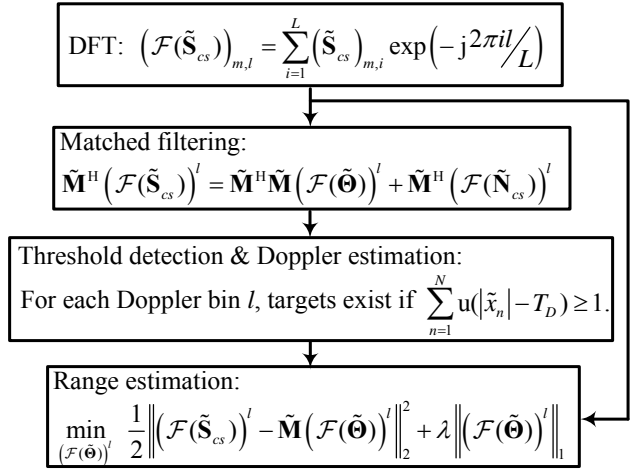

Fig.5: The mathematical procedure of the CoSaPD processing.

The block schematic of processing steps that are involved in the CoSaPD processing is given in Fig.4. The mathematical procedure corresponding to the processing blocks is given in Fig.5. It should be noted that the CoSaPD scheme firstly performs the estimation of the target velocities and then the estimation of the ranges. The operations are not reversible. The details of the detection process are depicted in the next section.

We now briefly discuss Doppler estimation in clutter. Different from thermal noise, the clutter has non-white power spectrum which is affected by radar and scenario parameters [1, 2, 26]. For stationary transceiver, clutter spectrum is around zero Doppler frequency. With this property, as in classic pulse-Doppler processing, the CoSaPD scheme can isolate the clutter from the moving target. If the target is away from the clutter spectrum, only thermal noise will interfere with its detection and then, each Doppler spectrum sample in the no-clutter area can be detected without clutter interference. If the target is in the clutter dominated area, the target power is not enough to perform detection. Then the Doppler spectrum samples in the area are simply discarded. However, because of clutter sidelobe, there is some clutter power at all Doppler frequencies, even though the clutter power is very small at high Doppler frequencies. To reduce the effects, we can add a data window [33] to weight the slow-time data for each virtual range bin, prior to computing the DFT. With the windowed data, the clutter has little effects on the estimation of targets as simulated in Section VI.

## V. Threshold Detection and Its Performance

This section describes the threshold detection used in the proposed CoSaPD processing and analyzes its performance.

Consider the $l$-th column or Doppler bin data derived from (20)

$$\left(\mathcal{F}(\tilde{\mathbf{S}}_{cs})\right)^l = \tilde{\mathbf{M}}\left(\mathcal{F}(\tilde{\mathbf{\Theta}})\right)^l + \left(\mathcal{F}(\tilde{\mathbf{N}}_{cs})\right)^l \tag{21}$$

Our purpose is to detect if there exists (at least) a target in the $l$-th Doppler bin. That is to determine if the vector $\left(\mathcal{F}(\tilde{\boldsymbol{\Theta}})\right)^l$ is a zero vector through detecting $\left(\mathcal{F}(\tilde{\mathbf{S}}_{cs})\right)^l$.

First we assume that the received noise is known. With the data (21), to further enhance the detection performance, we can perform the matched filtering of (21) as

$$\tilde{\mathbf{M}}^{\mathrm{H}}\left(\mathcal{F}(\tilde{\mathbf{S}}_{cs})\right)^l = \tilde{\mathbf{M}}^{\mathrm{H}}\tilde{\mathbf{M}}\left(\mathcal{F}(\tilde{\boldsymbol{\Theta}})\right)^l + \tilde{\mathbf{M}}^{\mathrm{H}}\left(\mathcal{F}(\tilde{\mathbf{N}}_{cs})\right)^l \tag{22}$$

which corresponds to the operation "Matched filtering" in the Fig. 4. For simplicity, let us define $\tilde{\mathbf{x}} = \tilde{\mathbf{M}}^{\mathrm{H}}\left(\mathcal{F}(\tilde{\mathbf{S}}_{cs})\right)^l$, $\tilde{\mathbf{y}} = \tilde{\mathbf{M}}^{\mathrm{H}}\tilde{\mathbf{M}}\left(\mathcal{F}(\tilde{\boldsymbol{\Theta}})\right)^l$ and $\tilde{\mathbf{w}} = \tilde{\mathbf{M}}^{\mathrm{H}}\left(\mathcal{F}(\tilde{\mathbf{N}}_{cs})\right)^l$. Then (22) is simply represented as

$$\tilde{\mathbf{x}} = \tilde{\mathbf{y}} + \tilde{\mathbf{w}} \tag{23}$$

Note that the noise term $\tilde{\mathbf{w}}$ in (23) is Gaussian but not independently distributed, because of the matched filtering. As illustrated in Section III, the matrix $\tilde{\mathbf{M}}$ is approximately column-by-column orthogonal. Then we can still assume that $\tilde{\mathbf{w}}$ is an *i.i.d.* Gaussian process. For the matched filtering data (22), the $k$-th target peak power is $4L^2|\rho_k|^2 T_b^2 B_{cs}^4 / B^2$ and noise variance is $4LN_0 T_b B_{cs}^3 / B$. Then the SNR after the matched filter from the $k$-th target, $\mathrm{SNR}_k^{MF}$, becomes to $T_b B_{cs} L$ times the received $\mathrm{SNR}_k^{IN}$.

The detection problem is to detect the targets from data $\tilde{\mathbf{x}}$. The binary detection problem of each $\tilde{x}_n$ ($1 \leq n \leq N$) can be formulated as

$$\begin{aligned} H_0 &: |\tilde{x}_n| = |\tilde{w}_n| \\ H_1 &: |\tilde{x}_n| = |\tilde{y}_n + \tilde{w}_n| \end{aligned} \tag{24}$$

The detection probability and false alarm probability are given respectively by

$$P_D^n = \int_{T_D}^{\infty} f_{|\tilde{x}_n\||H_1}(|\tilde{x}_n|\,|\,H_1) d|\tilde{x}_n| \tag{25}$$

$$P_F^n = \int_{T_D}^{\infty} f_{|\tilde{x}_n\||H_0}(|\tilde{x}_n|\,|\,H_0) d|\tilde{x}_n| \tag{26}$$

where $T_D$ is the detection threshold and $f_{|\tilde{x}_n\||H_1}(|\tilde{x}_n|\,|\,H_1)$ and $f_{|\tilde{x}_n\||H_0}(|\tilde{x}_n|\,|\,H_0)$ are

respectively the probability density functions of $|\tilde{x}_n|$ given that a target is present and not present. Then the false alarm probability and the detection probability of $\tilde{y}$ are given by

$$P_F \approx 1 - \prod_{n=1}^{N}(1-P_F^n) \tag{27}$$

$$P_D \approx 1 - \prod_{n=1}^{N}(1-P_D^n) \tag{28}$$

Define $\sigma^2 = 2LN_0 T_b B_{cs}^3 / B$. The $|\tilde{w}_n|$ has a Rayleigh density with mean $\sqrt{\pi/2}\sigma$ and variance $(4-\pi)\sigma^2/2$. Under hypothesis $H_0$, the target is absent, the pdf of $|\tilde{x}_n|$ is given by

$$f_{|\tilde{x}_n||H_0}(|\tilde{x}_n|\,|\,H_0) = \frac{|\tilde{x}_n|}{\sigma^2}\exp(-\frac{|\tilde{x}_n|^2}{2\sigma^2}) \tag{29}$$

Under hypothesis $H_1$, the target is present, $\tilde{x}_n = \tilde{y}_n + \tilde{w}_n$ is complex Gaussian distributed with mean $\tilde{y}_n$ and variance $2\sigma^2$. Then the pdf of $|\tilde{x}_n|$ follows Rician distribution,

$$f_{|\tilde{x}_n||H_1}(|\tilde{x}_n|\,|\,H_1) = \frac{|\tilde{x}_n|}{\sigma^2}\exp(-\frac{|\tilde{x}_n|^2+|\tilde{y}_n|^2}{2\sigma^2})\mathrm{I}_0(\frac{|\tilde{x}_n||\tilde{y}_n|}{\sigma^2}) \tag{30}$$

where $\mathrm{I}_0(\cdot)$ is the modified Bessel function of the first kind [34].

With the known noise power $\sigma^2$, the Neyman-Pearson optimal detector can be derived from the likelihood ratio test

$$\frac{f_{|\tilde{x}_n||H_1}(|\tilde{x}_n|\,|\,H_1)}{f_{|\tilde{x}_n||H_0}(|\tilde{x}_n|\,|\,H_0)} = \exp(-\frac{|\tilde{y}_n|^2}{2\sigma^2})\mathrm{I}_0(\frac{|\tilde{x}_n||\tilde{y}_n|}{\sigma^2}) \underset{H_0}{\overset{H_1}{\gtrless}} \lambda \tag{31}$$

For the monotonically increasing function $\mathrm{I}_0(\cdot)$, we have equivalent and simple expression of (31) as

$$|\tilde{x}_n| \underset{H_0}{\overset{H_1}{\gtrless}} \lambda_0 \sigma = T_D \qquad (32)$$

where $\lambda_0 = \sqrt{-2\ln(P_F^n)}$ is the scale factor used to control the false alarm rate. Thus we can derive the joint detector for a vector data $\tilde{\mathbf{x}}$ as

$$\sum_{n=1}^{N} \mathrm{u}\left(|\tilde{x}_n| - \lambda_0 \sigma\right) \underset{H_0}{\overset{H_1}{\gtrless}} 1 \qquad (33)$$

where $\mathrm{u}(\cdot)$ represents unit step function. The detection process for a Doppler bin is shown in Fig.6.

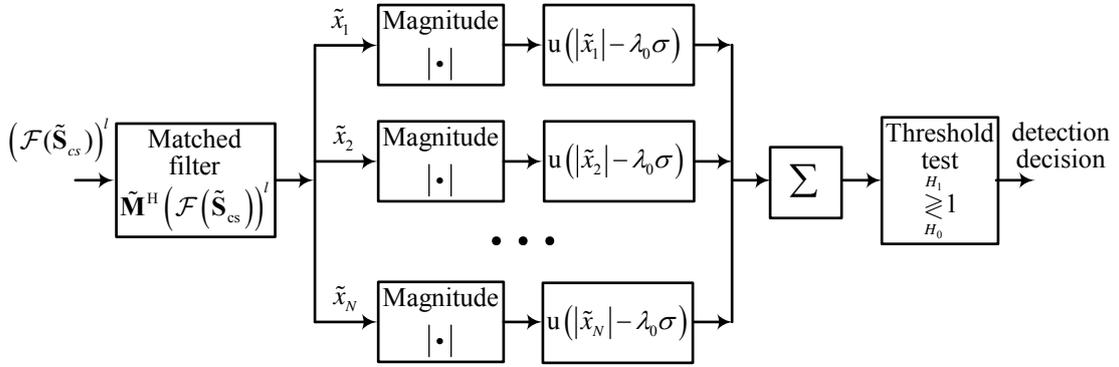

Fig.6: The block schematic of the detection process for a Doppler bin.

With (26), (29) and (32), we can derive $P_F^n = \exp(-\lambda_0^2/2)$. Then we have the false alarm probability as

$$P_F \approx 1 - \left(1 - \exp(-\lambda_0^2/2)\right)^N \qquad (34)$$

We are not able to derive the closed form of the detection probability $P_D$. However, it is noted that the CoSaPD detector can achieve the processing gain of $T_b B_{cs} L$, which is less than the gain of classic processing, $TB_b L$. Then it is expected that the performance of the proposed detector will degrade in a comparison with the classic detector at the low SNR.

After the target detection, the CoSaPD processing conducts the range estimation through the sparse recovery algorithms, as discussed in last Section. As is well-known, an inherent characteristic of the recovery algorithms [35, 36] is to detect the non-zero elements in the sparse vector. Then to improve the system detection performance, we may set low detection threshold for each Doppler bin. The low threshold will increase the false alarm probability and thus may introduce false targets. However, the detected false targets can be removed through the detection process in the recovery algorithms. The recovery algorithm strives for a minimum number of non-zero cells at its output. From the system point of view, the false alarm probability of the radar system will not increase, although we set the low detection threshold. The observations are confirmed in the next simulations.

In practice, it is impossible to know the noise parameter $\sigma$ in advance. To make the false alarm rate constant, we should estimate the parameter $\sigma$ to obtain an adaptive threshold. Following the assumptions on the measurement matrix $\tilde{\mathbf{M}}$, it is seen that the noise matrix $\tilde{\mathbf{M}}^H \mathcal{F}(\tilde{\mathbf{N}}_{cs})$ is independent and identically distributed and its element absolute $\left|\left(\tilde{\mathbf{M}}^H \mathcal{F}(\tilde{\mathbf{N}}_{cs})\right)_{i,j}\right|$ ($1 \leq i \leq N$, $1 \leq j \leq L$) follows the Rayleigh distribution. Then the maximum likelihood estimation of $\sigma$ is just the average of the available data [37],

$$\hat{\sigma} = \sqrt{\frac{2}{\pi}} \frac{\sum_{(i,j) \in \Lambda} \left|\left(\tilde{\mathbf{M}}^H \mathcal{F}(\tilde{\mathbf{N}}_{cs})\right)_{i,j}\right|}{|\Lambda|} \qquad (35)$$

where $\Lambda$ is the set consisting of all available $i$ and $j$, and $|\Lambda|$ is the size of $\Lambda$. For sparse targets, the accumulative strength of signals $\sum_{(i,j) \in \Lambda} \left|\left(\tilde{\mathbf{M}}^H \tilde{\mathbf{M}} \mathcal{F}(\tilde{\mathbf{\Theta}})\right)_{i,j}\right|$ is much smaller than that of noise $\sum_{(i,j) \in \Lambda} \left|\left(\tilde{\mathbf{M}}^H \mathcal{F}(\tilde{\mathbf{N}}_{cs})\right)_{i,j}\right|$, and then the following approximation is appropriate when $|\Lambda|$ is large

$$\frac{\sum_{(i,j)\in\Lambda}\left|\left(\tilde{\mathbf{M}}^{\mathrm{H}}\mathcal{F}(\tilde{\mathbf{N}}_{cs})\right)_{i,j}\right|}{|\Lambda|} \approx \frac{\sum_{(i,j)\in\Lambda}\left|\left(\tilde{\mathbf{M}}^{\mathrm{H}}\mathcal{F}(\tilde{\mathbf{S}}_{cs})\right)_{i,j}\right|}{|\Lambda|} \tag{36}$$

In the simulation study, we set $|\Lambda| = NL$ and have the estimated $\hat{\sigma}$ as

$$\hat{\sigma} \approx \sqrt{\frac{2}{\pi}} \frac{\sum_{i=1}^{N}\sum_{j=1}^{L}\left|\left(\tilde{\mathbf{M}}^{\mathrm{H}}\mathcal{F}(\tilde{\mathbf{S}}_{cs})\right)_{i,j}\right|}{NL} \tag{37}$$

Then the detection threshold is given as $T_D = \lambda_0 \hat{\sigma}$.

## VI. Simulations

In this section we present simulation performance of the proposed CoSaPD processing and make a comparison with classic processing [1,2] and direct processing by (19). Subsection A introduces simulation scenarios. Subsections B and C simulate detection and estimation performance in an additive white Gaussian noise. The effects of clutter are simulated in Subsection D.

### A. Simulation scenarios

It is assumed that the radar transmits a linear frequency modulation pulse train with carrier frequency $f_c = 10\text{GHz}$, signal bandwidth $B = 200\text{MHz}$, pulse width $T_b = 10^{-5}\text{s}$, pulse repetition interval $T = 10^{-4}\text{s}$. The coherent processing interval consists of $L = 100$ pulses. For the assumed parameters, the unambiguous target ranges and Doppler frequencies are in $1500\text{m} \sim 3466.5\text{m}$ and $-5\text{KHz} \sim 5\text{KHz}$, respectively. The range resolution is $0.75\text{m}$ and the Doppler resolution is $0.1\text{KHz}$.

For the QuadCS system, the chipping sequence $p(t)$ is generated by random $\pm 1's$ with rate $1/B$ and the bandpass filter is set to be an ideal one with bandwidth $B_{cs}$. Two bandpass filters with $B_{cs} = 50\text{MHz}$ and $B_{cs} = 25\text{MHz}$ are considered. For the two filters, the sampling rates in the low-rate samples are one fourth and one eighth the Nyquist rate, respectively. The basis pursuit denoising (BPDN) algorithm [28] is used for the sparse target recovery.

The simulated radar signal has a flat spectrum. The QuadCS measurement matrix $\tilde{\mathbf{M}}$ is approximately column-by-column orthogonal. Fig.7 shows the distribution of the averaged Gram matrix $\tilde{\mathbf{M}}^H\tilde{\mathbf{M}}$ over 1000 independent trials for $B_{cs} = 25\text{MHz}$. The maximum off-diagonal element of the Gram matrix is 0.015. The Gram matrix clearly demonstrates that the assumption is reasonable.

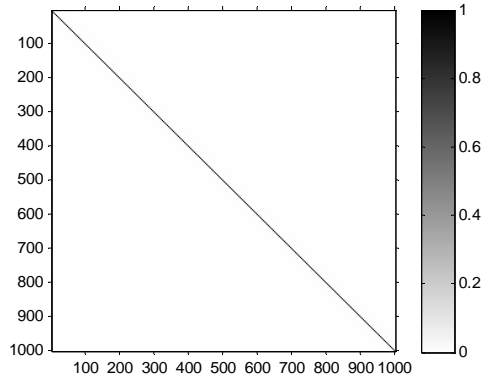

Fig.7: The distribution of the averaged Gram matrix $\tilde{\mathbf{M}}^H\tilde{\mathbf{M}}$.

## B. Detection performance

We assume that there are three targets with the same signal-to-noise ratio. The target delays and the Doppler frequencies are randomly set in the unambiguous region. We present three simulation results. For the first two results, the delays and the Doppler frequencies are in resolution grids. For the third result, the delays and the Doppler frequencies are arbitrarily set. To test the detectability of the multiple targets, the Doppler frequencies are set in the same Doppler bin. All results are obtained by averaging 1000 independent trials.

Firstly, we show that the CoSaPD detector is of the constant false alarm rate in a Doppler bin. Fig.8 shows the variations of the false alarm probabilities versus the scale factors for different signal-to-noise ratios when $B_{cs} = 25\text{MHz}$. It is seen that the change of the noise powers does not have effect on the false alarm probabilities for a specified scale factor, which is consistent with the theoretical result in (34). The same conclusion can be drawn for $B_{cs} = 50\text{MHz}$.

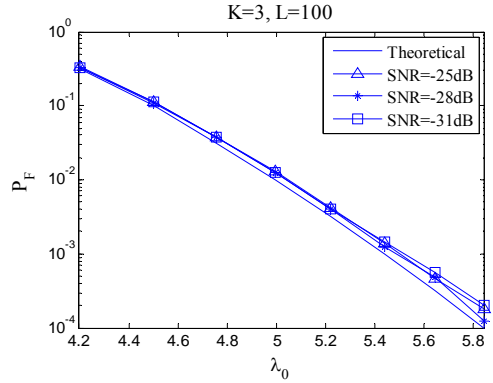

Fig.8: The false alarm probability versus the scale factor

Next, we simulate the receiver operating characteristic (ROC) of the CoSaPD detector. Fig.9 shows the averaged ROC curves with a comparison with classic processing. It is seen that the performance of the CoSaPD detector is not as good as that of the classic detection. The performance degradation of the proposed detector is due to the decrease of the SNR gains. In Nyquist-rate case, after the matched filtering and DFT processing, the processing gain can achieve $T_b B L$. While in the QuadCS system, it only realizes a gain of $T_b B_{cs} L$. As the bandpass width $B_{cs}$ increases, the processing gain $T_b B_{cs} L$ increases and then the detection performance gets better. In the simulated example, the processing gains achieve 53dB, 47dB and 44dB for classic processing and CoSaPD detector with $B_{cs} = 50\text{MHz}$ and $B_{cs} = 25\text{MHz}$, respectively. The SNRs for detection are 23dB, 17dB and 14dB, respectively. There are reductions of 6dB and 9dB in SNR relative to the classic processing for the 50MHz and 25MHz cases, respectively. For $B_{cs}=50\text{MHz}$ or the compressive sampling rate equal to one fourth the Nyquist rate, the detection performance of the CoSaPD detector almost achieves that of the classic detector in the simulated range of $P_F$.

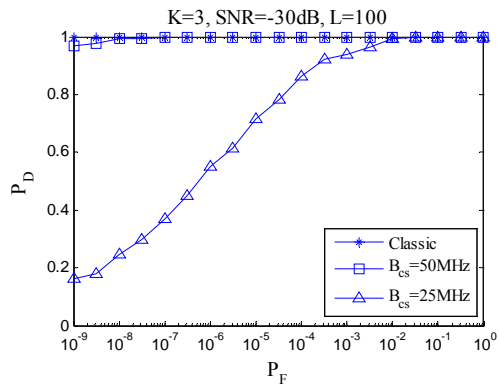

Fig.9: ROC of the CoSaPD detector.

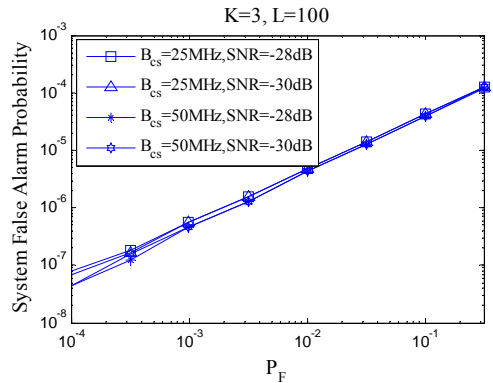

Fig.10: System false alarm probability versus detector false alarm probability.

As discussed in last Section and noted in Fig.9, we can set high false alarm probability to increase the detection probability. However, simply doing so will increase false targets. The problem can be resolved in the recovery stage of the target range following the detector. This is because the sparse recovery algorithm has the inherent detection ability [35, 36]. Fig.10 shows the false alarm rate of the system after the recovery stage versus that of the detector. Although the detector has high false alarm probability in the detection stage, the recovery algorithm can keep low false alarm probability of the system. Processing the system detection in this way will slightly increase the computational burden in the range estimation.

Fig.11 further shows the changes of detection performance as SNR varies when $P_F = 10^{-2}$. It can be seen that even at SNR=-30dB, the CoSaPD detector can approach to the performance in classic detection with one eighth the Nyquist rate.

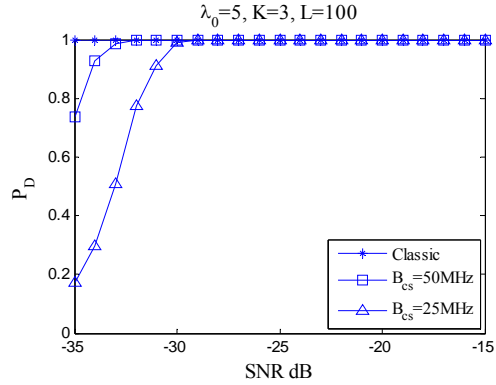

Fig.11: The detection performance versus SNR.

Finally, we consider a realistic scenario in which the ranges and Doppler frequencies of the targets are continuous and may not fall on the resolution grids. Fig.12 shows the ROC in this case. In comparison with Fig.9, the detection performance degrades. It is noted that when the target is not on Doppler bin, the target energy for the detection is from the Doppler leakage which is smaller than that of the target on the Doppler bin. When the target is away from the range bin, the measurement matrix contains errors which will degenerate the matched filter in (22). Then the detection performance is poorer than that of the targets on the resolution grids.

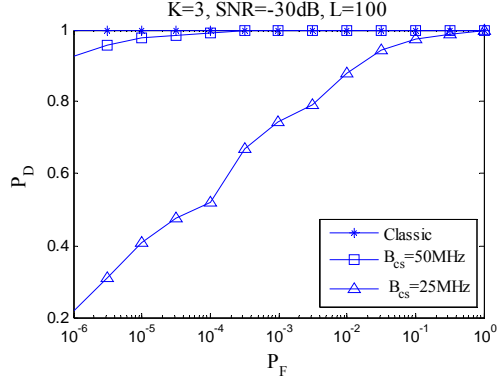

Fig.12: ROC in realistic case.

## C. Estimation performance

After detecting the existence of the targets in a specific Doppler bin, we perform the estimation of the corresponding target ranges. The CoSaPD method is depicted in the last stage of Fig.5 and is realized through the recovery algorithms. The rate of successful estimation is used as the performance metric. When the targets are at the resolution grids, a successful estimation refers to that the ranges and Doppler frequencies are correctly estimated; when the target ranges and Doppler frequencies are chosen randomly at the unambiguous region, a successful estimation is declared if the difference between the estimated and real ranges and/or Doppler frequencies is in half resolution cell.

In the simulation study, we set five targets with the same SNRs. To discuss the ability of discrimination, we assume that the first two targets are at the same range bin, the other two targets are at the same Doppler bin, the fifth target is randomly set for its range and Doppler. All results are obtained by averaging 1000 independent trials.

Firstly, we depict the estimation performance when the five targets are set at the discrete grids. Fig.13 shows the rates of successful estimation at different SNRs. For the CoSaPD method, the false alarm probability is set as $P_F = 10^{-2}$ with $\lambda_0 = 5$. It is seen that the CoSaPD method greatly outperforms the direct method and can achieve the performance of the classic method even at -25dB for one eighth the Nyquist rate. The performance improvement of the CoSaPD method is due to that the range estimation is performed in Doppler domain. The SNR of Doppler domain data is enhanced because of the DFT processing. Setting $P_F = 10^{-2}$ will result in high false alarm probability in detection stage. However, the setting does not affect the system detection.

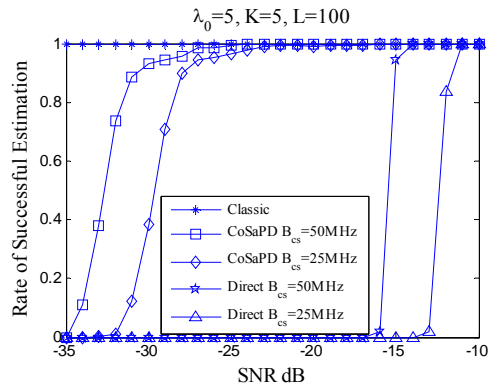

Fig.13: Rates of successful estimation for different methods.

Another advantage of the CoSaPD method over the direct method is the reduction of the computational burden. For the simulated parameters, the direct method needs to call 100 times recovery algorithms while the CoSaPD method only needs at most 12 times recovery algorithms, as shown in Fig.14.

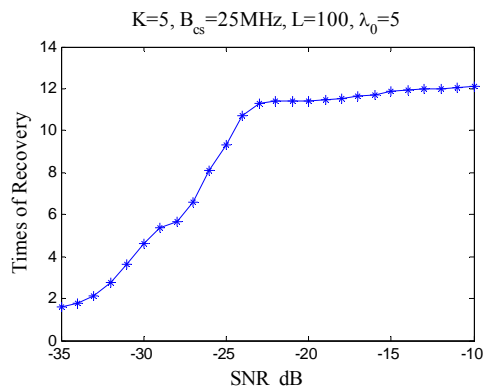

Fig.14: Times of the recovery algorithms used for range estimation.

Next, we present the simulation performance when the target ranges and Doppler frequencies are randomly set in the unambiguous region. Fig.15 shows the rates of successful estimation. In comparison with Fig.13, the estimation performance degrades. However, the CoSaPD method is more applicable to the realistic case than the direct method. As noted in Section IV, the direct method first estimates the complex amplitudes of the targets from the compressive data. The estimation may introduce errors in amplitude and phase. In particular, the phase error will greatly affect the Doppler estimation in DFT operation. Then the direct method is much poorer in the estimation performance. For the CoSaPD method, the Doppler estimation is performed in Doppler domain data which is from DFT of the compressive data. The range estimation by recovery algorithm contains no Doppler phase. Thus the CoSaPD method is robust in the practical case.

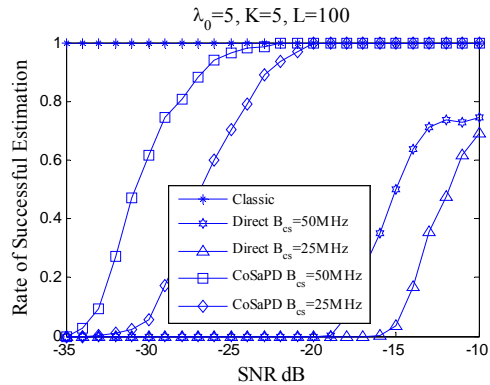

Fig.15: The rate of successful estimation in practical case.

Finally, we simulate the performance for estimating a smaller target around a stronger target. In the classic processing, the output of the matched filtering will have side lobes in range, which makes the smaller target barely visible above the side lobes of the stronger target. We assume that the two targets are in the same Doppler bin and the smaller target is randomly set in the first side lobe of the stronger target. Fig.16 shows the estimation performance, where $SNR_1^{IN}$ and $SNR_2^{IN}$ denote the SNRs of the stronger target and the smaller target, respectively. It is seen that the CoSaPD method outperforms the classic method when the two targets have large SNR difference.

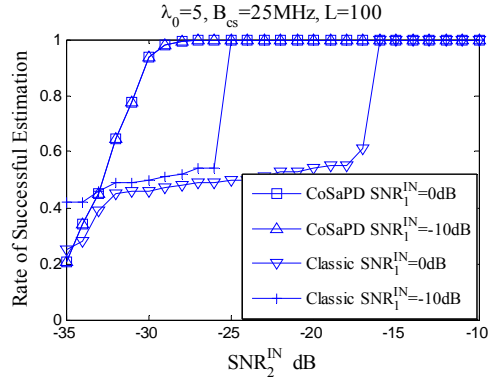

Fig.16: Estimation performance of the smaller target.

### D. The Effects of Clutter

We now demonstrate the performance of the CoSaPD processing in surface clutter. The received target signals are contaminated in both noise and clutter as in (5). The signals and noise are simulated as in Fig.13. The surface clutter is assumed to be Rayleigh-distributed in amplitude and obey the two-sided exponential law in Doppler spreading

$$S_c(v) = \frac{\beta}{2}\exp(-\beta|v|), \ -\infty < v < \infty$$

The Doppler model well describes windblown ground clutter with $\beta$ corresponding to the wind conditions [26]. In the simulation, $\beta = 4.3$, a wind condition about 60 miles per hour. The SCR is -40dB. To reduce the effect of clutter, a Taylor window with 10 nearly constant-level sidelobes adjacent to the mainlobe and a peak sidelobe level of -70 dB relative to the mainlobe peak is used before DFT processing.

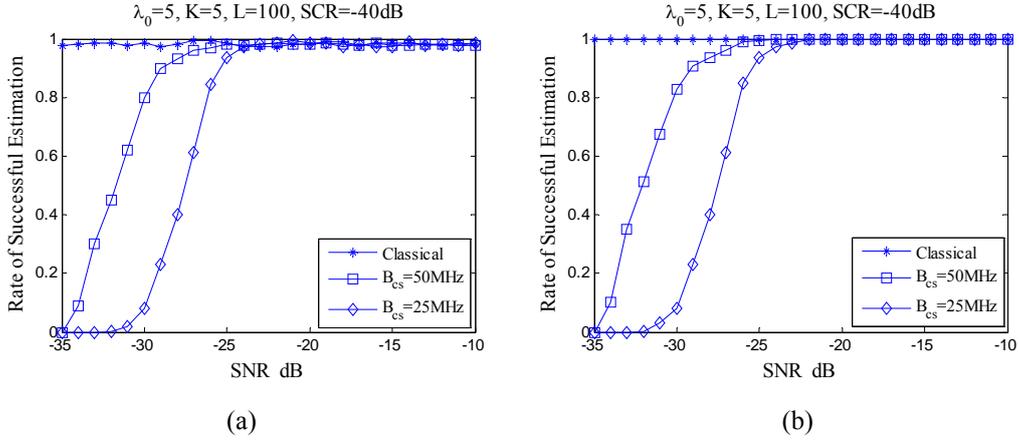

(a)  (b)

Fig.17: Rates of successful estimation in clutter for discarding
5 Doppler bins (a) and 13 Doppler bins (b).

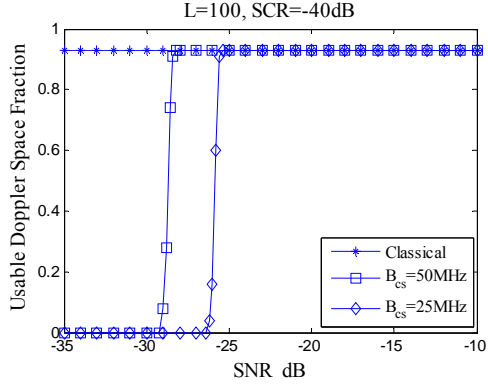

Fig.18: Usable Doppler space fraction versus SNR.

As discussed in Section IV, the Doppler spectrum samples will be discarded if the targets are in the clutter dominated area. Then we evaluate the rates of successful estimation at different SNRs for discarding different Doppler bins around zero Doppler shift. Fig.17 shows simulation results for discarding 5 and 13 Doppler bins, respectively. The effect of clutter is clear from Fig.17 (a). Because of clutter sidelobe, we cannot obtain 100% successful estimation rates both for the classic and CoSaPD methods. If we discard more Doppler bins, as shown in Fig.17 (b), we can almost completely remove the effect of the clutter on the signals with high Doppler shifts.

Fig.18 further shows usable Doppler space fraction [2] versus SNRs. In the simulation, we set one target with randomly distributed range bin and variable Doppler bin. When the target is successfully estimated, we claim that the Doppler bin is usable. It is seen that there is sharp drop of the usable Doppler space when the SNR is lower than some threshold. This is because the recovery algorithms are not workable in such low SNRs[2]. Both Fig.17 (b) and Fig.18 indicate that the CoSaPD method can achieve the performance of the classic method even at -25dB for one eighth the Nyquist rate. This is consistent with the observation from Fig.13.

## VII. Conclusion

This paper has developed a pulse-Doppler processing scheme, CoSaPD, with the sub-Nyquist data delivered from the QuadCS system. Owing to the data structure parallel to the classic sampling, the CoSaPD takes some ideas from the classic processing. The scheme follows the procedure of Doppler estimation/detection and range estimation, which is irreversible in the processing order. Theoretical analyses and computer simulations show its performance advantages. When sampling at one eighth the Nyquist rate and for SNR above -25dB, the CoSaPD achieves the performance of the classic processing with Nyquist samples.

In comparison with other related schemes utilizing CS data, the CoSaPD scheme has four distinct characteristics. One is small size dictionary. In other CS-based radar data processing, the dictionary is often 2-dimensional by discretizing both radar range and Doppler [21]. The CoSaPD scheme adopts 1-dimensional dictionary by only discretizing the radar range. The second one is the combination of estimation and detection processes. The combination has two advantages over separate estimation and detection: the improvement of detection performance and the cut-down of computational burden. The third characteristic is the ability to detect the smaller target around the stronger target. The last one is the ability to cancel the clutter echoes as in classic processing.

---

[2] The $l_1$-norm minimization algorithms in noise case can derive the sparse solutions only when $\left\|\left(\mathcal{F}(\tilde{\mathbf{N}}_{cs})\right)^l\right\|_2^2 < \left\|\left(\mathcal{F}(\tilde{\mathbf{S}}_{cs})\right)^l\right\|_2^2$ [38].

## Acknowledgement

The authors would like to thank the editor and the anonymous reviewers for their constructive suggestions and remarks, which brought to our attention several important issues and definitely improved the presentation and the technical scope of the paper.

This work was supported in part by the National Science Foundation of China under Grant 61171166 and 61101193.

A new quadrature sampling and processing approach.

*IEEE Transactions on Aerospace and Electronic Systems*, **25**, 5 (Sep. 1989), 733-748.

[5] Vaughan, R. G., Scott, N. L. and White, D. R.

The theory of bandpass sampling.

*IEEE Transactions on Signal Processing*, **39**, 9 (Sep. 1991), 1973-1984.

[6] Donoho, D. L.

Compressed sensing.

*IEEE Transactions on Information Theory*, **52**, 4 (Apr. 2006), 1289-1306.

[7] Baraniuk, R. G.

Compressive sensing.

*IEEE Signal Processing Magazine*, **24**, 4 (Jul. 2007), 118-121.

[8] Candes, E. J., Romberg, J and Tao, T.

Robust uncertainty principles: exact signal reconstruction from highly incomplete frequency information.

*IEEE Transactions on Information Theory*, **52**, 2 (Feb. 2006), 489-509.

[9] Laska, J., Kirolos, S., Massoud, Y., Baraniuk, R., Gilbert, A., Lwen, M. and Strauss, M.

Random sampling for analog-to-information conversion of wideband signals.

*IEEE Dallas Workshop on Design, Applications, Integration and Software*, Oct. 2006, pp. 119-122.

[10] Laska, J. N., Kirolos, S., Duarte, M. F., Ragheb, T. S., Baraniuk, R. G. and Massoud, Y.

Theory and implementation of an analog-to-information converter using random demodulation.

*IEEE International Symposium on Circuits and Systems*, May 27-30, 2007, pp. 1959-1962.

[11] Juhwan Yoo, Turnes, C., Nakamura, E. B., Le, C. K., Becker, S., Sovero, E. A., Wakin, M. B., Grant, M. C., Romberg, J., Emami-Neyestanak, A. and Candes, E.

A compressed sensing parameter extraction platform for radar pulse signal acquisition.

*IEEE Journal on Emerging and Selected Topics in Circuits and Systems*, **2**, 3 (Sep. 2012), 626-638.

[12] Taheri, O. and Vorobyov, S. A.
34